\documentclass{iau}
\usepackage{graphicx}

\title[CALIFA survey: Star formation histories as a function of time and space] 
{CALIFA survey: The spatially resolved star formation history of massive galaxies}

\author[The CALIFA collaboration]   
{Rosa  Gonz\'alez Delgado$^1$, Enrique P\'erez$^1$, Roberto Cid Fernandes$^{1,2}$, Rub\'en Garc\'{\i}a-Benito$^1$,  Andr\'e L. de Amorim$^2$, Sebasti\'an F.  S\'anchez$^1$, Bernd Husemann$^3$, Rafael L\'opez Fern\'andez$^1$, Clara Cortijo-Ferrero$^1$, Eduardo Lacerda$^2$, Damian Mast$^1$ and the CALIFA collaboration}

\affiliation{$^1$Instituto de Astrof\'{\i}sica de Andaluc\'\i a, Granada, Spain \\[\affilskip]
$^2$Universidade Federal de Santa Catarina, Florian\'opolis, Brazil \\[\affilskip]
$^3$ Leibniz-Institut f\"ur Astrophysik, Postdam, Germany}

\pubyear{2013}
\volume{295}  
\pagerange{xx--xx}
\jname{The intriguing life of massive galaxies}
\editors{D. Thomas, A. Pasquali \& I. Ferreras, eds.}
\begin{document}

\maketitle

\begin{abstract}
The Calar Alto Legacy Integral Field Area (CALIFA) is an ongoing 3D spectroscopic survey of 600 nearby galaxies of all kinds. This pioneer survey is providing valuable clues on how galaxies form and evolve. Processed through spectral synthesis techniques, CALIFA datacubes allow us to, for the first time, spatially resolve the star formation history of galaxies spread across the color-magnitude diagram. The richness of this approach is already evident from the results obtained for the first $\sim 1/6$ of the sample. Here we show how the different galactic spatial sub-components ("bulge" and "disk") grow their stellar mass over time. We explore the results stacking galaxies in mass bins, finding that, except at the lowest masses, galaxies grow inside-out, and that the growth rate depends on a galaxy's mass. The growth rate of inner and outer regions differ maximally at intermediate masses. We also find a good correlation between the age radial gradient and the stellar mass density, suggesting that the local density is a main driver of galaxy evolution.
\end{abstract}

\firstsection 
\section{Introduction}

Much of what we learnt about galaxies in the past decade has come from imaging or spectroscopic mega-surveys. Integral Field Spectroscopy (IFS) combines the best of these two age old observational techniques, adding the astrophysical diagnostic power of spectroscopy to the morphological information provided by imaging, so large scale IFS surveys are a natural next-step in extra-galactic research. CALIFA (S\'anchez et al 2012) is a pioneer in this blooming field. CALIFA is mapping the stellar populations, ionized gas and their kinematics for a representative sample of 600 galaxies in the nearby Universe ($0.005 < z < 0.03$), providing 3700--7000 \AA\ spectra for galaxies chosen to cover the color-magnitude diagram down to M$_r \leq -18$ mag  while filling the $\sim 1$ arcmin$^2$ field of view of the PPaK IFU 
mounted on the 3.5m at Calar Alto.

This contribution reports some of the initial results of our stellar population analysis of CALIFA datacubes. Following the footsteps of our own previous work with galaxy spectra, we retrieve physical properties and star-formation histories from $\lambda$-by-$\lambda$ spectral fits using the {\sc starlight} code, now applied in a spaxel-by-spaxel basis. The example from spatially-unresolved spectra from the SDSS illustrates both the complexities and the benefits of the combination of large-sample statistics with the manifold of galaxy properties derived from spectral analysis. With over 1000 spectra per galaxy, CALIFA rivals in quantity with the SDSS. Qualitatively, however, it is a completely different survey, one which looks {\em within} galaxies to find out what comes from where, as opposed to the global perspective obtained from integrated surveys.

Details on the observations and data reduction are given in Husemann et al.\ (2012). The processing of the datacubes through {\sc starlight} is discussed in Cid Fernandes et al.\ (2013). In what follows we highlight some of the results obtained from 107 galaxies observed as of March 2012 (see also P\'erez et al. 2013 and Gonz\'alez Delgado et al.\ 2013 for details and more results). Of the many possible approaches, and in keeping with the spirit of this meeting, we follow the strategy of presenting the radial variations of physical properties and star formation histories sorting galaxies by their stellar mass.


\section{Physical properties and star formation histories as a function of radius and mass}

\begin{figure}[t]
\begin{center}
 \includegraphics[width=0.7\textwidth,angle=-90]{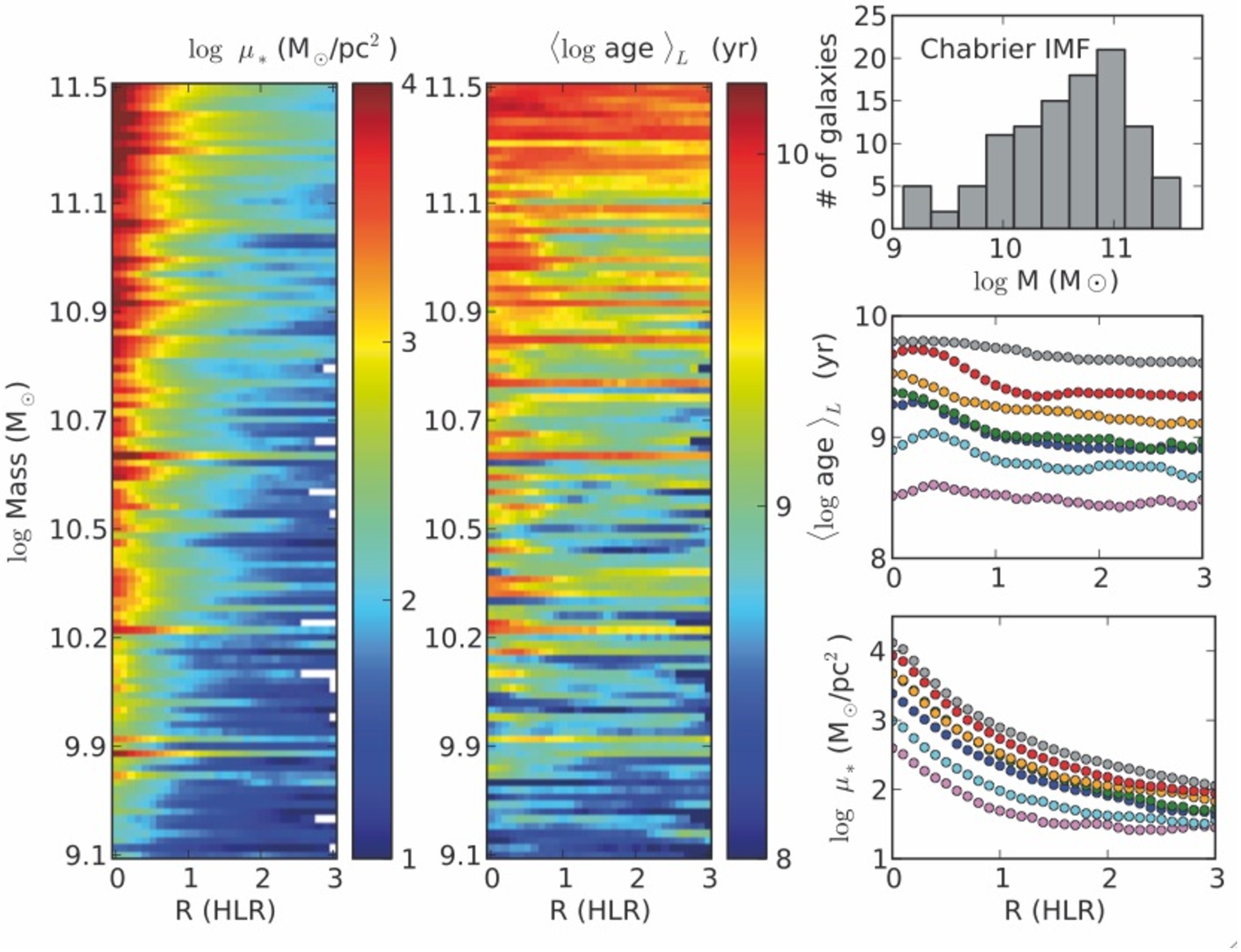} 
 \caption{{\em Left:} Surface stellar mass density radial profiles for 107 galaxies, as derived from the application of {\sc starlight} to the datacubes. Galaxies are vertically sorted by their total  stellar mass. {\em Middle:} Radial distributions of the luminosity-weighted mean stellar age obtained from the spectral decomposition in terms of different SSPs. {\em Right:} $\mu_\star$ and $\langle \log {\rm age} \rangle_L$ profiles obtained by stacking  galaxies in the same 7 contiguous mass intervals labeled in the left axis of the previous panels ($10^{9.1-9.9}$ \ldots $10^{11.1-11.5}$), and the histogram of the total stellar masses.
  }
   \label{fig:Fig1}
\end{center}
\end{figure}

The science/art of full spectral synthesis of stellar populations has progressed substantially in the past decade. In fact, it was one of the main tools employed in the exploration of the fossil record of galaxy evolution encoded in SDSS spectra (e.g., Panter et al. 2005; Asari et al. 2007; Tojeiro et al.\ 2009). The {\sc starlight} code decomposes galaxy spectra as combinations of simple stellar populations (SSP) of different ages and metallicities, providing estimates of stellar masses, mean ages and metallicities, kinematics, extinction, etc. With CALIFA, all these properties can now be mapped in $xy$ coordinates. The results shown below were obtained ensuring (by means of spatial binning, when necessary) a minimum $S/N$ ratio of 20 in the continuum at 5635 \AA, and exploring different sets of SSP models, including Bruzual \& Charlot (2003), Charlot \& Bruzual (2007, priv.\ comm.), Gonz\'alez Delgado et al.\ (2005) and Vazdekis et al.\ (2010).

Fig.\ \ref{fig:Fig1} shows azimuthally averaged radial profiles of the derived stellar mass density and the luminosity weighted mean log age. The radial coordinate is expressed in units of the galaxy's Half Light Radius (measured around the rest-frame V band) to facilitate the comparison. Note that the datacubes reach 2 to 4 HLR.
The plot shows how more massive galaxies have higher and more centrally concentrated stellar mass densities. The trend of older ages for more massive galaxies is also clearly visible, and at any $R$.

\begin{figure}[t]
\begin{center}
\includegraphics[width=3.0in,angle=-90]{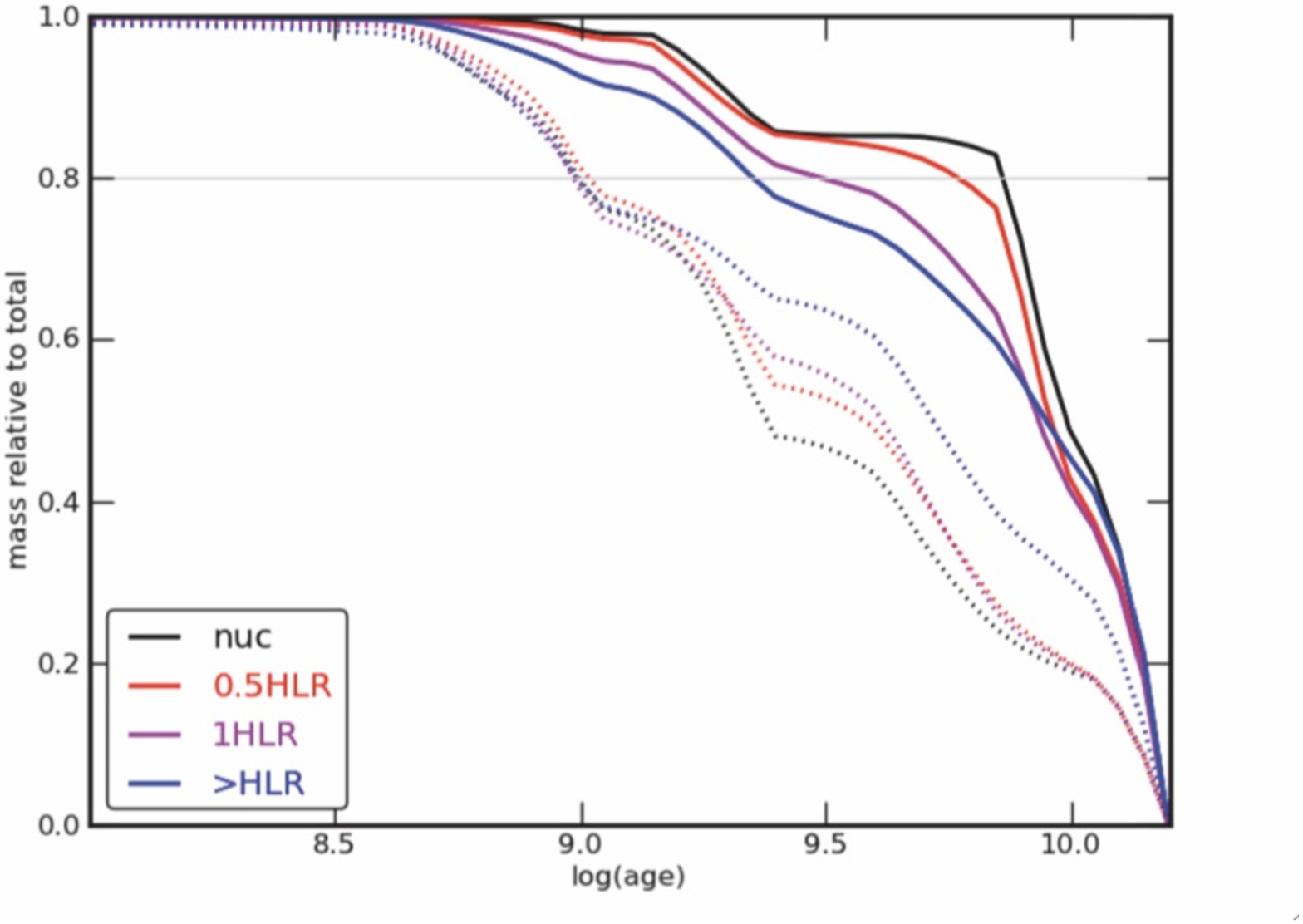} 
 \caption{Average mass growth as a function of lookback time (in yr) for galaxies in two mass ranges: $< 10^{10}$ (dotted lines) and 5--$7 \times 10^{10} M_\odot$ (solid). Different colors correspond to different radial zones, from the nucleus (black), $R < 0.5$  (red), $R < 1$ (purple)  to $R > 1$ HLR (blue).}
   \label{fig:Fig2}
\end{center}
\end{figure}

Perhaps the main appeal of spectral synthesis is its ability to recover (at least approximately) the actual time-dependent history of galaxies. With CALIFA we can now do this radius-by-radius. As an illustration, Fig.\ \ref{fig:Fig2} shows how the stellar mass grows in time, normalized such that the mass assembled up to today equals 1. Dotted and solid lines represent average curves for galaxies with masses in the $< 10^{10}$ and 5--$7 \times 10^{10} M_\odot$ ranges, respectively. Different spatial locations (defined in terms of each galaxy's own HLR) are coded by different colors. Again, the signature of downsizing is clear. Clearly, while at the lowest mass galaxies outskirts grow faster (dotted blue line steeper), at intermediate and high masses galaxies grow inside-out (full black line steeper).

\begin{figure}[t]
\begin{center}
\includegraphics[width=0.65\textwidth,angle=-90]{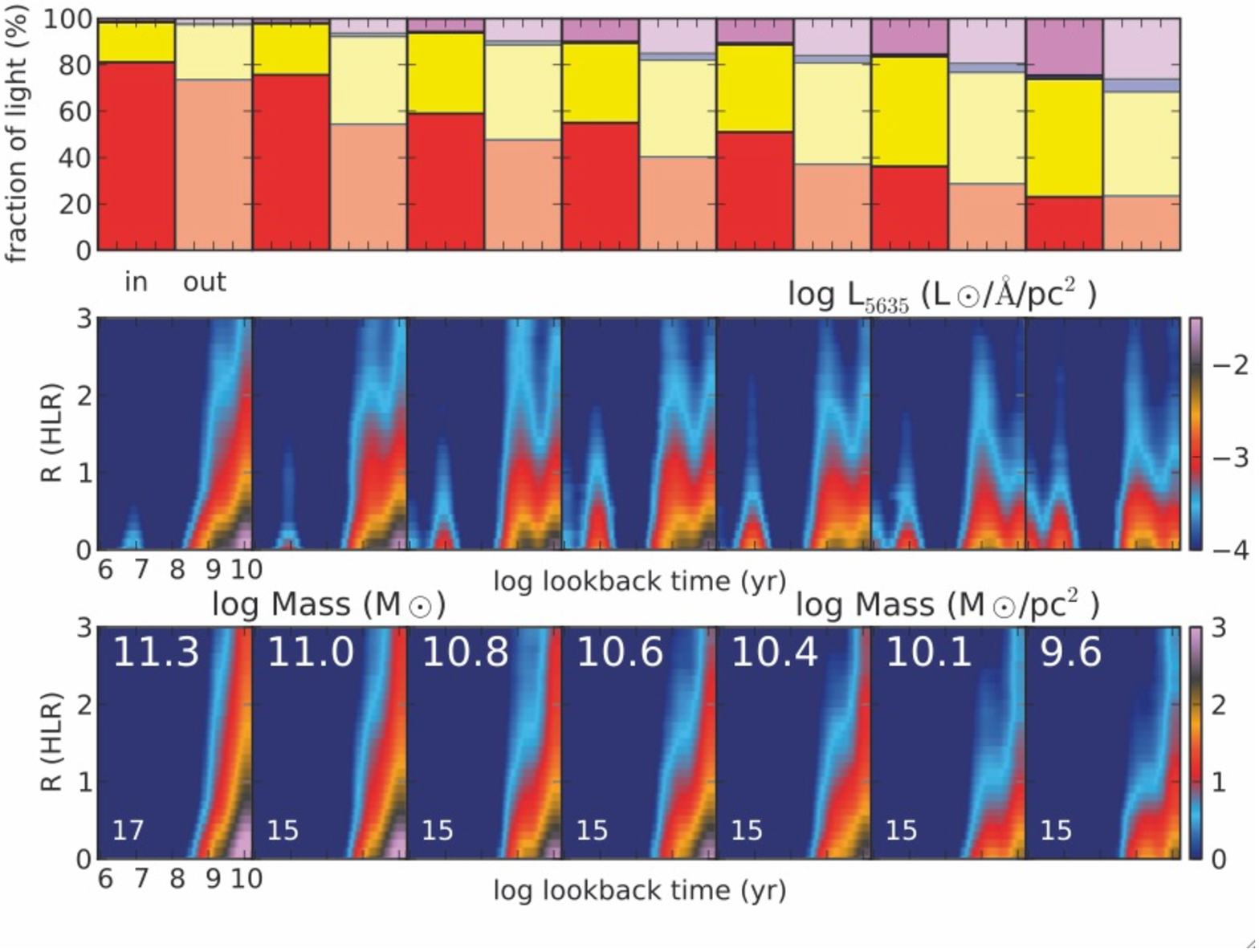} 
 \caption{Star formation history as a function of space and time, stacked in seven bins in galaxy mass (indicated in the bottom panels). {\em Top:} Fractional distribution of the optical light into populations of different age ranges: $> 2.5$ Gyr (red), 0.2--2.5 Gyr (yellow), 25--200 Myr (blue), $\leq 25$ Myr (magenta). Each panel contains one frame for the inner 0.5 HLR (left) and another for $R = 1$--2 HLR (right). {\em Middle \& bottom:}  $R$-$t$ diagrams showing the luminosity at $\lambda =$ 5635 \AA\ per unit area and stellar mass formed per unit area, resolved in space and time.} 
   \label{fig:Fig3}
\end{center}
\end{figure}

Fig.\ \ref{fig:Fig3} portrays the distribution of stellar light and mass as 
a function of both time and radius, for each of seven equally populated total mass bins.  Galaxies become progressively younger as their masses decrease. Note how the spatial-gradient in age (seen as the difference between inner and outer regions) is maximal for galaxies of intermediate mass. For both the most and the least massive galaxies, age gradients are very mild, as previously seen in Fig.\ \ref{fig:Fig1}. The middle panels of Fig.\ \ref{fig:Fig3} present a more detailed account of how the light and mass are distributed in space and (lookback) time. 

By opening the spatial dimension collapsed in previous surveys, CALIFA is also opening new ways to look at galaxies. Interpreting the results squeezed above (and the many others not even mentioned here) poses challenges to both theorists and observers. In the context of current ideas, intermediate mass galaxies are less susceptible to SN and AGN negative feedback, resulting in more efficient star-formation and a peak in the proportion of stellar to halo mass. This mass range coincides with that where we see the maximum inside $\times$ outside relative growth efficiency. This looks more like a clue than a coincidence. For more tips on the life of galaxies, keep an eye on upcoming CALIFA papers.

\vspace{0.2 cm} {\footnotesize 
\noindent
{\bf Acknowledgments:} This study makes use of the data provided by the Calar Alto Legacy Integral Field Area (CALIFA) survey (http://califa.caha.es/), the first legacy survey performed at Calar Alto. The CALIFA collaboration thanks the IAA-CSIC and MPIA-MPG as major partners of the observatory, and CAHA itself, for the unique access to telescope time and support in manpower and infrastructures. We thank  support of Spanish Ministerio de Econom\'{\i}a y Competividad through grant AYA2010-15080.}


\begin{thebibliography}{}

\bibitem[\protect\citeauthoryear{Asari}{2007}]{Asari07}  Asari, N. V. et al.\ 2007, MNRAS, 381, 263

\bibitem[\protect\citeauthoryear{Bruzual \& Charlot}{2003}]{BC03} Bruzual G., Charlot S., 2003, MNRAS, 344, 1000

\bibitem[\protect\citeauthoryear{Gonz\'alez Delgado}{2005}]{GD05} Gonz\' alez Delgado, R. M. et al.\ 2005, MNRAS, 357, 945

\bibitem[Panter et al.(2007)]{2007MNRAS.378.1550P} Panter, B., Jimenez, R., 
Heavens, A.~F., \& Charlot, S.\ 2007, MNRAS, 378, 1550 

\bibitem[\protect\citeauthoryear{Sanchez, S.F.,  et al.}{2012}]{Sanchez2012} S\'anchez, S. F. et al.\  2012, A\&A, 538, 8

\bibitem[Tojeiro et al.(2007)]{2007MNRAS.381.1252T} Tojeiro, R., Heavens, 
A.~F., Jimenez, R., \& Panter, B.\ 2007, MNRAS, 381, 1252 
\bibitem[\protect\citeauthoryear{V\' azdekis, A.  et al.}{2010}]{Vazdekis2010} 

Vazdekis, A. et al.\ 2010, MNRAS, 404, 1639

\end{thebibliography}
\end{document}